\pdfoutput=1 
\documentclass{IEEEtran4PSCC}
\ifCLASSINFOpdf
   \usepackage[pdftex]{graphicx}
\else
   \usepackage[dvips]{graphicx}
\fi
\usepackage[cmex10]{amsmath}
\usepackage{physics}
\usepackage{amssymb}
\usepackage{amsfonts} 
\usepackage{graphicx}
\usepackage{subcaption}

\usepackage{algorithmic}

\usepackage{array}
\makeatletter
\let\old@ps@headings\ps@headings
\let\old@ps@IEEEtitlepagestyle\ps@IEEEtitlepagestyle
\def\psccfooter#1{%
    \def\ps@headings{%
        \old@ps@headings%
        \def\@oddfoot{\strut\hfill#1\hfill\strut}%
        \def\@evenfoot{\strut\hfill#1\hfill\strut}%
    }%
    \def\ps@IEEEtitlepagestyle{%
        \old@ps@IEEEtitlepagestyle%
        \def\@oddfoot{\strut\hfill#1\hfill\strut}%
        \def\@evenfoot{\strut\hfill#1\hfill\strut}%
    }%
    \ps@headings%
}
\makeatother

\psccfooter{%
        \parbox{\textwidth}{\hrulefill \\ \small{23rd Power Systems Computation Conference} \hfill \begin{minipage}{0.2\textwidth}\centering \vspace*{4pt} \includegraphics[scale=0.06]{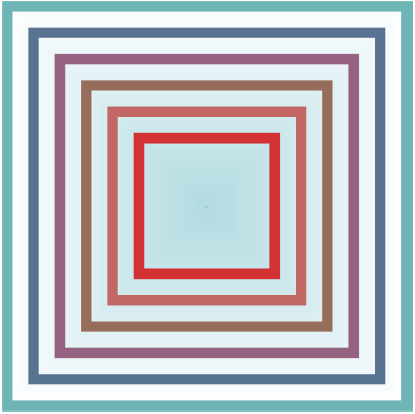}\\\small{PSCC 2024} \end{minipage} \hfill \small{Paris, France --- June 4 -- 7, 2024}}%
}

\begin{document}
%
\title{Quantum Amplitude Estimation for Probabilistic Methods in Power Systems}

\author{
\IEEEauthorblockN{Emilie Jong, Brynjar Sævarsson, Hjörtur Jóhannsson, Spyros Chatzivasileiadis}
\IEEEauthorblockA{Department of Wind and Energy Systems \\
Technical University of Denmark (DTU)\\
Kgs. Lyngby, Denmark\\
\{emijo, brysa, hjjo, spchatz\}@dtu.dk}
}


\maketitle

\begin{abstract}
This paper introduces quantum computing methods for Monte Carlo simulations in power systems which are expected to be exponentially faster than their classical computing counterparts. Monte Carlo simulations is a fundamental method, widely used in power systems to estimate key parameters of unknown probability distributions, such as the mean value, the standard deviation, or the value at risk. It is, however, very computationally intensive. Approaches based on Quantum Amplitude Estimation can offer a quadratic speedup, requiring orders of magnitude less samples to achieve the same accuracy. This paper explains three Quantum Amplitude Estimation methods to replace the Classical Monte Carlo method, namely the Iterative Quantum Amplitude Estimation (IQAE), Maximum Likelihood Amplitude Estimation (MLAE), and Faster Amplitude Estimation (FAE), and compares their performance for three different types of probability distributions for power systems.
\end{abstract}


\thanksto{\noindent Submitted to the 23rd Power Systems Computation Conference (PSCC 2024).}

\section{Introduction}
The introduction of renewable energy sources in the power system together with an increasingly fluctuating net electricity demand drastically increases the degree of stochasticity in power system operation. To ensure a secure and efficient operation of power systems, we need to estimate both accurately and efficiently probability distributions for wind forecast errors, line loadings, electricity demand patterns, voltage profiles, and many others. Key parameters, such as the mean value, standard deviation, and value at risk are important in order to extract information that will help assess the degree of risk and determine the optimal power system operation. Traditionally, Monte Carlo simulations have been widely used to estimate key parameters of probability distributions, such as the mean value or the value at risk. Considering, however, the high computational burden of such approaches, new computational algorithms are extremely important to help drastically reduce the computational time for these tasks. \newline
\indent Quantum Amplitude Estimation (QAE) algorithms, as first introduced in \cite{QAEBrassard}, have proven to show a quadratic speed-up compared to Classical Monte Carlo simulations (CMC). Although the initial algorithm \cite{QAEBrassard} requires extensive operations which make it unfeasible for current quantum computers due to Quantum Phase Estimation (QPE) (i.e. quantum computers belonging to the Noisy Intermediate-Scale Quantum era), in the previous years, variants on QAE have been developed requiring less gates and qubits and thus making them more suitable for implementation on near-term quantum computers. Among them, the most promising are the Iterative Quantum Amplitude Estimation (IQAE) \cite{Grinko_2021}, the Maximum Likelihood Amplitude Estimation (MLAE) \cite{QAEwithoutQPE}, and the Faster Amplitude Estimation (FAE) \cite{Nakaji_2020}. All algorithms require less samples to achieve the same accuracy and confidence level as Classical Monte-Carlo Simulations (CMC) and do not require the Quantum Phase Estimation. FAE was introduced as a faster variant that required even less samples than the first version of IQAE. Two years after its first introduction in 2019, IQAE was updated in 2021 \cite{Grinko_2021} with improved theoretical bounds, giving it a renewed theoretical advantage over FAE. When it comes to MLAE, Ref. \cite{yu2020comparison} compares the performance of IQAE and MLAE in terms of oracle queries and circuit depth for a given degree of accuracy. In this paper it is concluded that MLAE and IQAE are comparable in accuracy and number of oracle calls and predominate classical Monte Carlo integration. \newline
\indent The goal of this paper is to introduce the three most promising Quantum Amplitude Estimation algorithms for power systems applications by providing a condensed overview of the research conducted in \cite{emilie}, and compare them theoretically and empirically with regards to their efficiency and estimation error. We implement them using Qiskit \cite{qiskit}, the most widely used Python-based toolbox for Quantum Computing, and explore their quantum advantage compared to the Classical Monte Carlo Simulations. We apply the algorithms to different probability distributions for power system applications to estimate parameters such as the mean, the value at risk and conditional value at risk. \newline
\indent From a classical point of view, the required amount of samples for estimating the mean with a certain accuracy and confidence level is largely influenced by the type of probability distribution that it is applied on. Depending on the power system application, stochastic values follow different types of probability distributions. For example, wind forecasts typically follow a Weibull distribution, while wind forecast errors are often considered to follow a normal distribution with zero mean. Uniform distributions are also widely used, e.g. for the distribution of voltage magnitudes across system nodes. This paper introduces the three most promising quantum computing variants which have the potential to replace classical Monte Carlo simulations for power systems, and assesses them against Monte Carlo simulations for three different types of probability distributions: normal, Weibull, and uniform. We evaluate the performance of these algorithms based on the number of samples they need, their accuracy, estimation errors, and circuit depth. \newline
\indent The remainder of the paper is structured as follows: In Section \ref{sec:qcfundamentals}, we delve into the fundamental principles of quantum computing. Section \ref{sec:qae} introduces the Quantum Amplitude Estimation (QAE) algorithms and provides a theoretical comparison. Section \ref{sec:impl} describes the practical application and implementation of the algorithms. This is followed by the presentation of the results obtained from simulations on quantum computers in Section \ref{sec:results}, including a comparative analysis of the QAE algorithms. Moving forward to Section \ref{sec:disc}, we discuss the limitations in this research. Finally, we conclude in Section \ref{sec:concl}, summarizing the key findings and implications.


\section{Quantum Computing Fundamentals} \label{sec:qcfundamentals}
Classical bits process information using bits, which are either 0 or 1, while quantum bits (qubits) can be in a superposition state. The qubit's state remains uncertain until measured, at which point the qubit collapses into a definite state. By making use of this property, quantum computers can evaluate multiple states at the same time, allowing them to be more efficient than their classical counterparts. A qubit is represented by the Dirac bra-ket notation. A vector is denoted as $\ket{v}$, i.e. the \textbf{ket}. The \textbf{bra}, i.e $\bra{v}$,  is the conjugate transpose of $\ket{v}$. We can write the state of qubit $\ket{\psi}$ as:
\begin{equation} \label{eq:state}
    \ket \psi =  \alpha \ket 0 +  \beta \ket 1
\end{equation}

where $\alpha, \beta \in \mathbb{C}$, i.e they are complex numbers. 
$|\alpha|^2$ and $|\beta|^2$ are the probability of the qubit being in state $\ket 0$ and $\ket 1$ respectively, so it must hold that $|\alpha|^2 + |\beta|^2 = 1$. Quantum circuits make use of gates, which are operations that are applied to qubits to change their states. Important gates are the Hadamard gate, $H = \frac{1}{\sqrt{2}} \begin{bmatrix} 1 & 1 \\ 1 & -1 \end{bmatrix}$, and the Pauli matrices: \begin{align*} \label{eq:Pauligates}
        X = \begin{bmatrix}
            0 & 1 \\ 1 & 0
        \end{bmatrix}, && Y = \begin{bmatrix}
            0 & -i \\ i & 0 
        \end{bmatrix}, && Z = \begin{bmatrix}
            1 & 0 \\ 0 & -1
        \end{bmatrix}.
\end{align*} 
The more gates are used, the more the depth of the circuit increases. Although Quantum Computing has been rapidly evolving and the prospects for quantum computing applications look bright, we still find ourselves in the Noisy Intermediate-Scale Quantum (NISQ) era, where the circuit depth matters significantly due to the amount of noise generated by the gates. Meanwhile, publicly available quantum computers during this research were limited to only 7 qubits, while in October 2023 quantum computers with up to 127 qubits became publicly available for a time-limited use. 

\section{Quantum Amplitude Estimation} \label{sec:qae}
Quantum Amplitude Estimation is the quantum alternative of the Monte Carlo method. Theoretically, QAE shows a quadratic speed-up over classical algorithms \cite{quantumriskanalysis}. This means that instead of requiring 100'000 samples to estimate the mean value of a distribution, we would only need approx. 320 samples to estimate the mean with the same accuracy. In 2000, the initial canonical QAE version was invented, however it required a form of Quantum Phase Estimation (QPE) resulting in a very deep circuit. In the previous years, several other algorithms have been introduced, which are more suitable for quantum computers in the Noisy Intermediate-Scale Quantum (NISQ) era. This section introduces first the canonical version, and then the IQAE, MLAE and FAE algorithms. 
\subsection{Canonical Amplitude Estimation -- Proof of the Quadratic Speedup}
QAE was first introduced in \cite{QAEBrassard} using Grover's search algorithm and QPE. It works as follows. Suppose we have an operator $\mathcal{A}$ acting on $n+1$ qubits resulting in:
\begin{equation} \label{eq:canQAE}
    \ket{\psi} = \mathcal{A} \ket 0 _{n+1} = \sqrt{1-a}\ket{\psi_0}_n \ket 0 + \sqrt{a} \ket{\psi_1}_n \ket 1
\end{equation}
The application of $\mathcal{A}$ corresponds to an oracle query (quantum sample). Now, we can find an estimate for the amplitude of $a$ for state $\ket{\psi_1}$ using the Grover operator: $\mathcal{Q} = \mathcal{A}\mathcal{S}_0\mathcal{A}^\dagger \mathcal{S}_{\psi_0}$. $\mathcal{S}_0$ and $\mathcal{S}_{\psi_0}$ are reflections, defined as $\mathcal{S}_0 = \mathbb{I} - 2 \ket 0_{n+1} \bra{0}_{n+1}$ and $\mathcal{S}_{\psi_0} = \mathbb{I} - 2 \ket{\psi_0}_n \bra{\psi_0}_n \otimes \ket 0 \bra 0$ with $\mathbb{I}$ being the identity operator. For canonical QAE $m$ ancilla bits\footnote{Ancilla bits are auxiliary bits which are used as ``flags'' in quantum operations.} are needed to represent the final result, which are initialized in a superposition state created by Hadamard gates. 
The number of quantum samples is defined as $M = 2^m$. The ancilla bits control the increasing powers of $\mathcal{Q}$. Then a Quantum Fourier Transform (QFT) is applied to the ancilla bits, after which they are measured. The state of the ancilla bits corresponds to an integer $y \in \{0,..., M-1\}$ which is mapped to an angle $\theta_a = y\pi/M$. The estimate for $a$, $\tilde{a} = \sin^2{\theta_a} \in [0,1]$. Using this, \eqref{eq:canQAE} can be rewritten as with $0 \leq \theta_a \leq \pi/2$:
\begin{equation}
    \ket{\psi} = \mathcal{A} \ket{0}_{n+1} = \cos{\theta_a} \ket{\psi_0}_n \ket{0} + \sin{\theta_a} \ket{\psi_1}_n \ket 1
    \end{equation}
The estimate satisfies the bound error with a probability of $\frac{8}{\pi^2} \approx 81\%$:
\begin{equation} \label{eq:errorbound}
    \frac{2\sqrt{a(1-a)}\pi}{M} + \frac{\pi^2}{M^2} = \mathcal{O} \left ( \frac{1}{M} \right )
\end{equation}
This shows a quadratic speedup compared to classical Monte Carlo methods which converge with $\mathcal{O} \left ( \frac{1}{\sqrt{M}} \right )$ \cite{quantumriskanalysis}. In \cite{QAEBrassard}, it is shown that after $k$ applications of $\mathcal{Q}$ on $\ket \psi$, we get:
\begin{equation} \label{eq:Qm}
    \mathcal{Q}^k \ket \psi = \cos{((2k+1)\theta_a)} \ket{\psi_0} \ket{0} + \sin{((2k+1)\theta_a)} \ket{\psi_1} \ket{1}
\end{equation}

The canonical algorithm then uses QPE which requires many controlled $\mathcal{Q}$ operation and additional ancilla qubits. IQAE, MLAE and FAE differ from the canonical algorithm  by eliminating the need for quantum phase estimation through a combination of post-processing techniques and by applying different powers of the Grover operators.

\subsection{Iterative Quantum Amplitude Estimation}
In 2019, a new quantum amplitude estimation algorithm without phase estimation is proposed in \cite{Grinko_2021}, called Iterative Quantum Amplitude Estimation (IQAE). It uses Grover iterations to find an estimate for the target amplitude. IQAE starts with a small $\mathcal{Q}^0$ circuit which has one solution. This solution is then used to evaluate correct solutions from deeper circuits with more precise results (with higher powers of the Grover operator). It iteratively applies rotations to find the correct solution of the deeper circuits by using the outcomes of smaller circuits. It does this until it obtains an estimate with a certain confidence level. The advantage of IQAE is that it is possible to specify the desired confidence level. However, the circuit depth of IQAE cannot be controlled.
\subsection{Maximum Likelihood Amplitude Estimation} \label{subsec:MLAE}
The Maximum Likelihood Amplitude Estimation algorithm implements QAE without QPE, by using maximum likelihood as a post-processing method. Maximum likelihood is a statistical method used to estimate the parameters of a probability distribution by obtaining the values that would be the most probable when assuming that the observed data follows a certain distribution. In Ref. \cite{QAEwithoutQPE} it has been proven that the MLAE algorithm achieves nearly the optimal quantum speedup with a reasonable circuit length allowing MLAE to be executed on NISQ devices. MLAE runs experiments with multiple $\mathcal{Q}^k$ circuits. MLAE takes the likelihood functions of the results of these circuits and then applies the classical maximum likelihood estimation to retrieve an estimate for $\theta$. MLAE can thus be parallelized. There are two ways of finding the $k$ values for the power of the Grover operator: linearly and exponentially. The exponential approach results in a deeper circuit, however, it provides a more accurate estimation. The amount of query calls to achieve estimation error $\epsilon$ for MLAE is implied to be $\mathcal{O}\left (\frac{1}{\epsilon^{3/4}}\right )$ and $\mathcal{O}\left (\frac{1}{\epsilon}\right )$ for linearly and exponentially incremental sequence respectively compared to $\mathcal{O}\left (\frac{1}{\epsilon^{2}}\right )$ for the required samples of the classical approach in order to achieve the same precision. Because of the more accurate estimation of the exponentially incremental sequence, we will consider just the exponentially incremental sequence.

\subsection{Faster Amplitude Estimation}
Faster Amplitude Estimation (FAE) is proposed in \cite{Nakaji_2020}. This algorithm is similar to IQAE, however, it solves an ambiguity differently in the post-processing. For FAE, it is possible to fix the amount of iterations to control the circuit depth as opposed to IQAE.

\subsection{Theoretical Quantum Advantage}
In quantum, a sample is defined as the amount of times $\mathcal{A}$ is applied and is called an oracle query or oracle call. The estimation accuracy depends on the confidence level $\alpha$ and the estimation error $\epsilon$. The estimation error is the difference between the actual value and the estimated value of the parameter. The confidence level indicates the probability that the estimate has a maximal estimation error of $\epsilon$ and is defined as: 100(1-$\alpha$)\%. For example, if $\alpha$ is set to 0.05, $P[|\theta - \hat{\theta}| \leq \epsilon] \geq 95\%$, where $\hat{\theta}$ is the estimate of $\theta$.

For MLAE, the lower bound for the amount of samples $N$ is given by \eqref{eq:NqMLAE}. 
\begin{equation} \label{eq:NqMLAE}
    N_{min}^{MLAE} \geq \frac{\sqrt{\alpha(1-\alpha)}}{\epsilon}
\end{equation}

For the other algorithms, upper bounds are provided. For IQAE, different upper bounds are presented in \cite{Grinko_2021}. The loose upper bound presented in the paper is:
\begin{equation} \label{eq:Nq_IQAE}
    N_{max}^{IQAE} < \frac{50}{\epsilon}\log{\left (\frac{2}{\alpha}\log_2{\left ( \frac{\pi}{4 \epsilon}\right)} \right )}
\end{equation}
Another smaller upper bound is given for IQAE, which is based on an empirical complexity analysis: Clopper-Pearson (CP). The Clopper-Pearson bound is:
\begin{equation} \label{eq:Nq_avg_CP}
    N_{max}^{CP-avg} \leq  \frac{0.8}{\epsilon} \log{\left (\frac{2}{\alpha}\log_2{\left ( \frac{\pi}{4 \epsilon}\right)} \right )}
\end{equation}

Finally, the upper bound for FAE is:
\begin{equation} \label{eq:Nq_FAE}
    N_{max}^{FAE} <  \frac{4.1\cdot 10^3}{\epsilon} \log{\left (\frac{2}{\alpha}\log_2{\left ( \frac{2\pi}{3 \epsilon}\right)} \right )}
\end{equation} 
For classical Monte Carlo, the relationship between the accuracy and the amount of samples follows the central limit theorem:
\begin{equation} \label{eq:NqCMC}
    N_{min}^{CMC} \sim \frac{z_c(\alpha)^2 s_n^2}{\epsilon^2}, 
\end{equation}
where the value of $z_c$ depends on the confidence level $\alpha$ and $s_n$ is the sample standard deviation of the distribution. 

\begin{figure}
    \centering
    \includegraphics[width=0.5\textwidth]{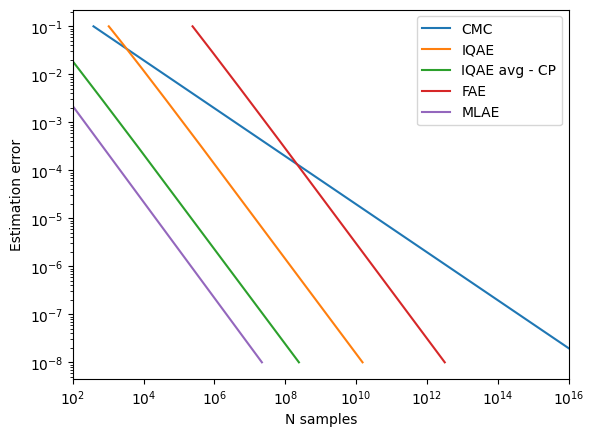}
    \caption{Comparison of the theoretical speedup of QAE algorithms and the classical Monte Carlo approach ($\alpha = 0.05$) \cite{emilie}}
    \label{fig:comparisonQAE}
\end{figure}

Fig.~\ref{fig:comparisonQAE} shows a graphical comparison of the algorithms. Here, for CMC, $z_c = 1.96$ for a confidence interval of 95\% ($\alpha = 0.05$) and $s_n^2 = 1$, but in reality $s_n$ depends on the sampling and on the type of probability distribution, e.g. the more complex and asymmetric the distribution, the larger $s_n$. The graph depicted for CMC is thus more an indication for its trend, since this graph shifts upwards or downwards depending on the $s_n$. It should be noted that for all algorithms, except MLAE and CMC, upper bounds are shown, while for MLAE a lower bound is visualized.
\section{Implementation} \label{sec:impl}
\subsection{Method for estimating the mean, VaR and CVaR}
The mean value, the Value-at-Risk (VaR) and the Conditional-Value-at-Risk (CVaR) are the statistical parameters of interest for the investigated probability distributions, because they provide a measure for risk assessment. The VaR is a measure that estimates the maximum value in a system for a specified confidence level, while the CVaR measures the expected value if the VaR threshold is exceeded. The methods as explained in \cite{IBMcreditrisk} are used to evaluate the mean, VaR and CVaR. In order to be able to estimate these parameters, the qubits need to be initialized so that they represent the probability distribution. Suppose we have a random variable $X$ that follows a certain probability distribution. The probability distribution has to  be discretized into $2^n$ bins so that it can be represented by $n$ qubits in the interval [0, ..., N - 1], where $N = 2^n$. The discrete probability distribution has to be loaded into the amplitudes of the qubits. The quantum state for $X$ is then described as:
\begin{equation}
    \ket{\psi}_n = \sum_{x=0}^{N-1} \sqrt{p_x} \ket{x}_n,
\end{equation}

where $p_x$ is the probability of the state $\ket{x}_n$ and the sum of the probabilities of all states should be 1. This way, the sampling probability, which is the square of the amplitude, equals the probability of the distribution. The linear amplitude function $F$
implements piecewise linear functions on qubit amplitudes constructed by linearly controlled Pauli-Y rotations. The operator $F$ acts on an ancilla bit and maps the function $f \in [0, ..., 2^{n-1}]$ to:
\begin{equation} \label{eq:F}
    F|x\rangle|0\rangle = \sqrt{1 - f(x)} |x\rangle|0\rangle + \sqrt{f(x)}
    |x\rangle|1\rangle.
\end{equation}
Applying $F$ to $\ket{\psi_n}\ket{0}$ gives:
\begin{equation}
    \sum_{x=0}^{N-1} \sqrt{1-f(x)} \sqrt{p_x}\ket{x}_n \ket{0} + \sum_{x=0}^{N-1} \sqrt{f(x)} \sqrt{p_x}\ket{x}_n \ket{1}
\end{equation}
The probability of measuring $\ket{1}$ in the rightmost qubit is then equal to the expected value of $f(x)$. \newline
To estimate the value at risk, an operator, $\mathcal{F}_l$, is constructed which forms a cumulative distribution function. $F_l$ implements the function $f_l$ where $f_l(x) = 1$ for $x \leq l$ and $f_l(x) = 0$ otherwise. Applying $\mathcal{F}_l$ to $\ket{\psi}_n\ket{0}$ results in:
\begin{equation}
    \sum_{x=l+1}^{N-1} \sqrt{p_x} \ket{x} \ket{0} + \sum_{x=0}^{l} \sqrt{p_x} \ket{x} \ket{1}
\end{equation}
By using QAE, we can estimate the probability of obtaining $\ket{1}$ which is equal to  $\sum_{x=0}^{l}p_x = P[X\leq l]$. Then, classically, we perform a bisection search. The bisection search finds the smallest value $l_\alpha$ for which $P[X\leq l_\alpha]\geq 1- \alpha$. Quantum amplitude estimation is specifically suitable for estimating tail probabilities of distributions  \cite{IBMcreditrisk}. This holds, because for the VaR, we can replace $a$ in \eqref{eq:errorbound} for $\alpha$, since the estimated probability is larger than or equal to $\alpha$. If $\alpha$ is close to either 0 or 1, the error bound becomes very small. The error bound is then independent of other properties of the problem.

For estimating the CVaR, we identify the function $f(x) = \frac{x}{l_\alpha}f_l$. Applying the corresponding operator $\mathcal{F}$ yields:
\begin{multline} \label{eq:fcvar}
           F \ket{\psi}_n \ket{0} = \left ( \sum_{x = l_\alpha+1}^{N-1} \sqrt{p_x} \ket{x} + \sum_{x=0}^{l_\alpha} \sqrt{1-\frac{x}{l_\alpha}} \sqrt{p_x} \ket{x} \right ) \ket{0} \\ + \sum_{x=0}^{l_\alpha}\sqrt{\frac{x}{l_\alpha}} \sqrt{p_x}\ket{x} \ket{1} 
\end{multline} 
Then, we apply QAE to estimate the CVaR. Applying QAE gives $\sum_{x=0}^{l_\alpha}\frac{x}{l_\alpha}p_x$, where $\sum_{x=0}^{l_\alpha}p_x$ is $P[X\leq l_\alpha]$. Then, it follows that:
\begin{equation}
    CVaR(X) = \frac{l_\alpha}{P[X\leq l_\alpha]}\sum_{x=0}^{l_\alpha}\frac{x}{l_\alpha}p_x
\end{equation}

\subsection{Simulation procedure}
The QAE algorithms (IQAE, MLAE and FAE) are applied to the normal, Weibull and uniform distributions to find an estimate for the mean, VaR and CVaR. The parameters of the distributions are often based on historical data \cite{RAMADHANI2020106003}. Here, we will follow a more general approach for determining the parameters. Since the probability distribution is loaded on the amplitude of the qubits, the distribution can be scaled as desired by multiplying the qubit states accordingly. The normal distribution is denoted as $N(\mu, \sigma)$ with $\mu$ being the mean of the distribution and $\sigma$ the standard deviation. We investigate two normal distributions with the same mean ($\mu = 0.1$) and different standard deviations i.e. $\sigma = 0.01$ and $\sigma=0.05$. Normal distributions are applied in power systems to model uncertainty. We set the standard deviation to 10\% and 50\% as percentage of the mean value. In \cite{normal1}, for the uncertainty modelling of wind farms, the standard deviation can be up to 50\% of the mean value, while for load modelling the standard deviation is assumed to be 5\% of the mean
value. Another study fixes the standard deviation to 7\% of the mean value \cite{plfpvev}.
The Weibull distribution is referred to as $W(\beta)$, where $\beta$ is the shape parameter of the Weibull slope. We follow a general approach and just consider the shape parameter to model the shape of the distribution, while the scale parameter is set to 1. The shape parameter depends on the wind speed data, namely the average wind speed and the standard deviation \cite{fee0d0cf293b443f83a0ff7b4ec7c8ac}. 
We choose $\beta = 1.8$ for the Weibull distribution, since this parameter is typically between 1.63 to 2.97 \cite{weibullparameters}, determining the shape of the distribution. The uniform distribution is defined as $U(a, b)$ where $U$ is uniform on the interval $[a, b)$. The interval that is investigated is $[0, 1)$. The probability distribution is loaded onto 4 qubits. There is a trade-off between accuracy and runtime, while taking the limited amount of qubits in publicly available quantum computers into account. \newline
\indent For the QAE algorithms, the functions `IterativeAmplitudeEstimation', `MaximumLikelihoodEstimation' and `FasterAmplitudeEstimation' are used, which are available in Qiskit. To account for the varying scaling of the different probability distributions, we assess the relative errors of the application of different QAE algorithms applied to the different probability distributions in comparison to the actual distribution parameters. All algorithms are simulated 10 times on the quantum computer simulator to find the mean, VaR and CVaR for each QAE algorithm for all probability distributions. Furthermore, some tests are ran on a real quantum computer of IBM (IBM Guadalupe). Then, a more detailed comparison of the QAE algorithms versus CMC is performed for the estimate of the mean. For the comparison of QAE algorithms when estimating the mean, the average relative estimation error is computed for a range of oracles from $\sim 10^{2}$ to $\sim 2\cdot 10^5$. In \cite{sun2018probabilisticOPF}, it is suggested that  10000 iterations would provide sufficiently accurate results for probabilistic (optimal) power flow problems, however, is stated to be computationally intensive. In \cite{PPFstochastic}, it is mentioned that  tens of thousands of deterministic power flow calculations are needed for convergence with random samples generated from the probability distributions of the input variables. For this reason, the range that is investigated, is from order of magnitude $10^{2}$ to $10^5$. Thereafter, the practical confidence interval is computed to evaluate the spread of the estimates.

\section{Simulation} \label{sec:results}
In Table \ref{tab:rel_error_mean}, the average amount of oracle queries is shown with the corresponding relative error for IQAE, MLAE and FAE when estimating the mean of a normal distribution $N(0.1, 0.01)$. Here, we can observe that the order of magnitude for the average error differs, as well as the number of oracle queries. 

\begin{table} [!ht]
    \renewcommand{\arraystretch}{1.3}
    \centering
    \caption{Oracle queries and relative error (\%) for estimating the mean of N($\mu = 0.1, \sigma = 0.01$) ($\alpha = 0.05$, $N_{shots} = 100$, $\epsilon = 10^{-3}$, $q = 4$, $max\_iter = 3$)}
    \begin{tabular}{|c|c|c|}
    \hline
        QAE algorithm & Average oracle queries & $\Delta_{avg}$ (\%) \\
        \hline
        IQAE & 15960 & $0.03132$ \\
        \hline
        MLAE & 700 & $0.70745$ \\
        \hline
        FAE & 136238 & $1.55281$\\
        \hline
    \end{tabular}
    \label{tab:rel_error_mean}
\end{table}

\begin{figure}[h]
    \centering
    \includegraphics[width=0.5\textwidth]{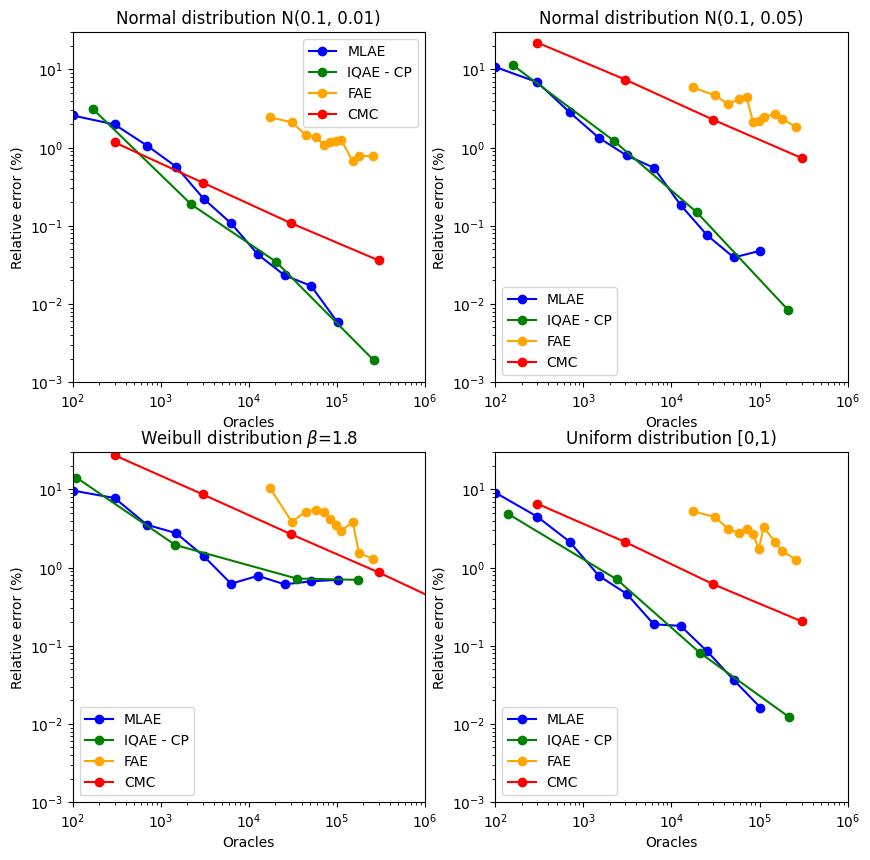}
    \caption{The relative error vs the amount of oracles plotted for estimating the mean of different types of probability distributions}
    \label{fig:empiricalestimationerror}
\end{figure}

\begin{figure}[!ht]
     \centering
     \begin{subfigure}[b]{0.4\textwidth}
         \centering
        \includegraphics[width=0.8\textwidth]{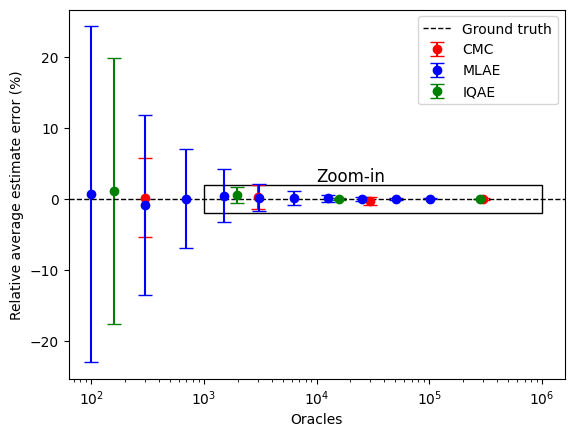}
         \caption{The relative deviation of the relative estimate from the actual mean and confidence interval for estimating the mean of N(0.1, 0.05)}
         \label{fig:avgconfint1}
     \end{subfigure}
     \hfill
     \begin{subfigure}[b]{0.4\textwidth}
         \centering
        \includegraphics[width=0.8\textwidth]{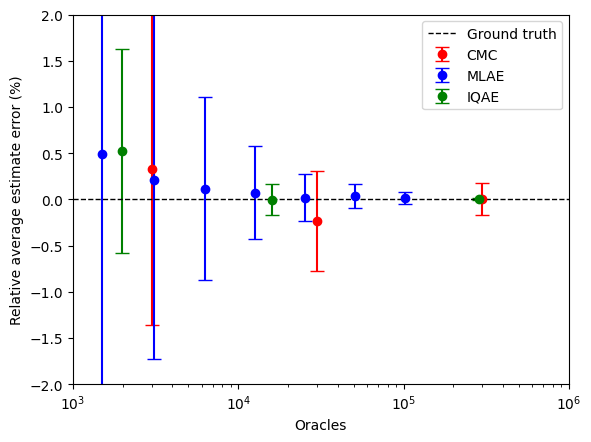}
         \caption{Zoom-in on the relative deviation of the relative estimate from the actual mean and confidence interval for estimating the mean of N(0.1, 0.05) for range $10^3-10^6$}
         \label{fig:avgconfint2}
     \end{subfigure}
     \caption{A zoom-in on the relative deviation of the relative estimate from the actual mean and confidence interval for estimating the mean of N(0.1, 0.05)}
     \label{fig:avgconfint}
\end{figure}

For a more consistent comparison of the algorithms, we assess the relative error for estimating the mean of the four probability distributions for each of the QAE algorithms and plot them using the same range of oracle queries. This is shown in Fig.~\ref{fig:empiricalestimationerror}. For a 1\% relative error when estimating the mean, the QAE algorithms do not seem to have any advantage over classical Monte Carlo in terms of oracle queries for the normal distribution N(0.1, 0.01). This could be the case because the standard deviation of the normal distribution is relatively small. Starting from a relative estimation error of 0.3\%, the IQAE and MLAE algorithms overtake CMC. Both algorithms show similar trends for all probability distributions regarding the average relative error and the amount of oracle queries. For N(0.1, 0.05) the quantum advantage for IQAE and MLAE is already clear for a relative error of around 10\%. The graph indicates that a smaller amount of oracles are needed when compared to the amount of samples needed for CMC to achieve the same accuracy. The advantage increases monotonically when we require smaller relative errors. Remarkably, the performance of FAE for N(0.1, 0.05) is much closer to CMC than for N(0.1, 0.01). 
Tables \ref{tab:var} and \ref{tab:cvar} show the results for estimating the VaR and the CVaR for the different distributions. The VaR levels deviate a few percentage points, since only 4 qubits are used for loading the probability distribution. Increasing the amount of qubits from 4 to 7 significantly decreases the relative error. For the uniform distribution, 4 qubits are not sufficient to find a VaR of 95\%. The average relative error for the CVaR is around 10 times larger than for the VaR. Again, increasing the amount of qubits for the loading of the probability distribution  to 7, significantly decreases the average relative error. \newline
\noindent From Fig.~\ref{fig:empiricalestimationerror}, it can be concluded that the IQAE and MLAE algorithms show similar performance in terms of accuracy and required amount of oracle calls. For further comparison, we investigate the confidence interval for the estimates that IQAE, MLAE and CMC provide. Fig.~\ref{fig:avgconfint} shows the average confidence interval for the estimates given by IQAE, MLAE and CMC. FAE has been left out of the graph for readibility reasons due to its significantly larger confidence interval compared to MLAE, IQAE and CMC. For samples in the order of magnitude of $10^2$, the confidence interval of the CMC is smaller than for the QAE algorithms. However, from $\sim10^3$ oracles, IQAE and MLAE overtake CMC. The confidence interval of IQAE is the smallest, and thus gives the most accurate estimate.
\begin{table}[!ht]
    \renewcommand{\arraystretch}{1.3}
    \centering
    \caption{Results for estimating the VaR for different distributions ($\alpha = 0.05$, $N_{shots} = 100$, $\epsilon = 10^{-3}$, $q = 4$, $max\_iter = 3$) including the actual VaR's}
    \label{tab:var}
    \begin{tabular}{|c|c|c|c|}
    \hline
         & IQAE & MLAE & FAE \\
        \hline
        \hline
        \multicolumn{4}{|c|}{\textbf{N (0.1, 0.01)}, $VaR_{95\%}= 0.1164$} \\
        \hline
        Oracles & 26840 & 700 & 121907 \\
        \hline
        VaR estimate & 0.1233 & 0.1233 & 0.1233 \\
        \hline
        VaR level (\%) & 97.92 & 98.77 & 97.92\\
        \hline
        Actual VaR for VaR level & 0.1204 & 0.1225 &  0.1204 \\
        \hline
        $\Delta_{VaR}$ (\%) & 2.41 & 0.65 & 2.41 \\
        \hline
        \hline
        \multicolumn{4}{|c|}{\textbf{N (0.1, 0.05)}, $VaR_{95\%} = 0.1822$} \\
        \hline
         Oracles & 23890 & 700 & 121907 \\
        \hline
        VaR estimate & 0.2100 & 0.2080 & 0.2040 \\
        \hline
        VaR level (\%) & 97.86 & 97.01 & 97.01 \\
        \hline
        Actual VaR for VaR level & 0.2013 & 0.1941 &  0.1941 \\
        \hline
        $\Delta_{VaR}$ (\%) & 4.14 & 7.16 &  5.10 \\
        \hline
        \hline
         \multicolumn{4}{|c|}{\textbf{Weibull ($\beta = 1.8$)}, $VaR_{95\%} = 1.8396$} \\
        \hline
         Oracles & 18030 & 700 & 121907 \\
        \hline
        VaR estimate & 2.0667 & 2.0893 & 2.0667 \\
        \hline
        VaR level (\%) & 96.52 & 96.66 & 96.50 \\
        \hline
        Actual VaR for VaR level & 1.9601 & 1.9734 &  1.9582 \\
        \hline
        $\Delta_{VaR}$ (\%) & 5.44 & 5.87 &  5.54 \\
        \hline
        \hline
        \multicolumn{4}{|c|}{\textbf{Uniform [0, 1)}, $VaR_{95\%} = 0.95$} \\
        \hline
         Oracles & 19890 & 700 & 121907 \\
        \hline
        VaR estimate & 1.0000 & 1.0000 & 1.0000 \\
        \hline
        VaR level (\%) & 100 & 100 & 100 \\
        \hline
    \end{tabular}
\end{table}
\begin{table}[!ht]
    \renewcommand{\arraystretch}{1.3}
    \centering
    \caption{Results for estimating the CVaR for different distributions ($\alpha = 0.05$, $N_{shots} = 100$, $\epsilon = 10^{-3}$, $q = 4$, $max\_iter = 3$) including actual CVaRs}
    \label{tab:cvar}
    \begin{tabular}{|c|c|c|c|}
    \hline
         & IQAE & MLAE & FAE \\
        \hline
        \hline
        \multicolumn{4}{|c|}{\textbf{N (0.1, 0.01)}} \\
        \hline
        Oracles & 18390 & 700 & 268924 \\
        \hline
        CVaR estimate & 0.1407 & 0.1390 & 0.1472 \\
        \hline
        VaR level (\%) & 97.92 & 98.77 & 97.92\\
        \hline
        Actual CVaR for VaR level & 0.1241 & 0.1259 & 0.1241 \\
        \hline
        $\Delta_{CVaR}$ (\%) & 13.40 & 16.85 & 12.08\\
        \hline
        \hline
        \multicolumn{4}{|c|}{\textbf{N (0.1, 0.05)}} \\
        \hline
         Oracles & 19420 & 700 & 280744 \\
        \hline
        CVaR estimate & 0.1875 & 0.1877 & 0.1873 \\
        \hline
        VaR level (\%) & 97.86 & 97.01 & 97.01 \\
        \hline
        Actual CVaR for VaR level & 0.2198 & 0.2135 &  0.2125 \\
        \hline
        $\Delta_{CVaR}$ (\%) & 14.69 & 12.06 &  11.86 \\
        \hline
        \hline
         \multicolumn{4}{|c|}{\textbf{Weibull ($\beta = 1.8$)}} \\
        \hline
         Oracles & 19470 & 700 & 277788 \\
        \hline
        CVaR estimate & 2.7305 & 2.7165 & 2.604 \\
        \hline
        VaR level (\%) & 96.52 & 96.66 & 96.50 \\
        \hline
        Actual CVaR for VaR level & 2.2528 & 2.651 &  2.2515 \\
        \hline
        $\Delta_{CVaR}$ (\%) & 21.21 & 19.93 &  15.64 \\
        \hline
    \end{tabular}
\end{table}
\begin{figure}[!ht]
     \centering
     \begin{subfigure}[b]{0.2\textwidth}
         \centering
         \includegraphics[width=\textwidth]{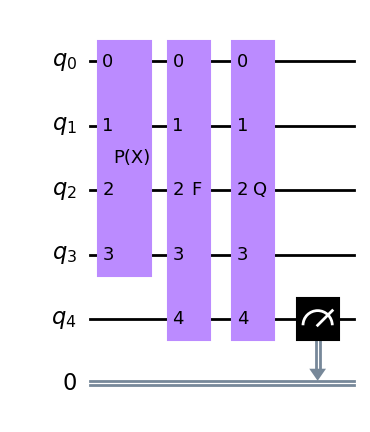}
         \caption{Estimating the mean using QAE with 1 iteration (k=1)}
         \label{fig:qae1iter}
     \end{subfigure}
     \hfill
     \begin{subfigure}[b]{0.3\textwidth}
         \centering
         \includegraphics[width=\textwidth]{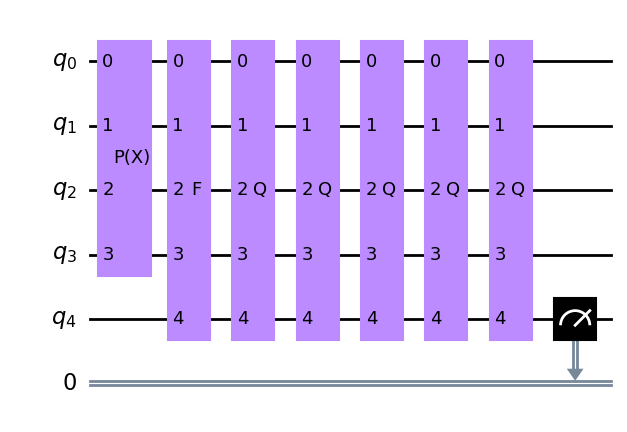}
         \caption{Estimating the mean using QAE with 5 iterations (k=5)}
         \label{fig:qae5iter}
     \end{subfigure}
     \caption{Circuits with different iterations for QAE for loading the probability distribution onto 4 qubits, $\mathcal{Q}$ is the Grover operator}
     \label{fig:qaeiter}
\end{figure}

The QAE algorithms have also been ran on a real IBM quantum computer (IBM Guadalupe) in order to estimate the mean value. However, due to the noise already generated in the gates of the circuit for the construction of the probability distribution, there are discrepancies in the representation of the probability distributions. This is why the CVaR and VaR have not been estimated on a real quantum computer. The construction of the probability distribution can already take up to 62 gates (in case of the Weibull distribution) and applying IQAE requires around 300 more gates. This does not yet allow for accurate results on a real quantum computer. 
\section{Discussion} \label{sec:disc}
The results show potential for the application of QAE to estimate the mean, VaR and CVaR for probability distributions relevant to power systems. The empirical comparison shows that MLAE and IQAE provide a speed-up compared to classical Monte Carlo simulation. For the tested probability distributions, the IQAE and MLAE algorithms were able to provide a more accurate estimate for the mean using less samples for all distributions. In case of N(0.1, 0.01), MLAE and IQAE gained an advantage over CMC around $\sim 3\cdot 10^3$ samples, while for the other distributions, this happened from $\sim 3 \cdot 10^2$ samples. FAE, however, did not show any advantage for the investigated range of samples. 

For the normal distribution and the uniform distribution similar trends are visible for the empirical and theoretical comparison while for the Weibull distribution at some point, the quantum advantage decreases. One of the reasons for this might be that the mean can not be estimated more accurately when more oracles are used due to the limited amount of qubits with which the probability distribution is represented. It could be that for the other distributions, the same happens when investigating an even larger range of oracles. It is assumed that increasing the amount of qubits would solve this problem.

\indent The accepted average relative error depends on the type of application for which they are used. Although IQAE seems to be the preferred algorithm in terms of number of oracle queries and accuracy, it should be noted that its circuit depth cannot be controlled, while for MLAE and FAE this is possible. For current quantum computers, the depth of a circuit still matters significantly due to the amount of noise generated by the quantum computers. The circuit depth increases drastically when the amount of Grover operations increases. For example, estimating the mean of $N(0.1, 0.01)$ using 1 iteration, the amount of required gates would be 299, while for 5 iterations, this number incresases to 1183. Fig.~\ref{fig:qaeiter} shows the circuits with 1 and 5 iterations respectively. Additionally, as the number of qubits within the circuit grows, the circuit depth increases as well.

\indent Although, IQAE and MLAE have proven to provide a speed-up compared to classical CMC empirically, it has to be noted that this advantage is still not visible in terms of runtime. While it takes several minutes to do one iteration of a QAE algorithm on the quantum computer simulator, it takes around 10-20 seconds to perform it on a quantum computer. However, CMC just needs several seconds in classical computation time. The advantage is thus purely seen in computation complexity.

\indent While in this paper, we have showed that we can apply QAE to known probability distributions, in a typical Monte Carlo simulation the probability distribution is not known a priori. The next step involves loading a truly unknown distribution on the qubits. 

\section{Conclusion} \label{sec:concl}
In this paper, we have introduced quantum computing algorithms for Monte Carlo simulations which have the potential to find wide use in power system applications. Monte Carlo simulations is a standard method employed in both power system literature and practice to measure the mean value or the value at risk of unknown distributions such as the wind forecast error or the line limit violations (among several others). Quantum Computing algorithms, based on the Quantum Amplitude Estimation method, have been theoretically proven to have a quadratic advantage versus classical Monte Carlo methods. For example, instead of requiring 100'000 samples to estimate the mean value of a distribution, they can use only approx. 320 samples to estimate it with the same level of accuracy. 

In this paper, we have described and tested the three most promising quantum algorithms that could potentially replace the Classical Monte Carlo simulations: the IQAE (Iterative Quantum Amplitude Estimation), the MLAE (Maximum Likelihood Amplitude Estimation), and the FAE (Faster Amplitude Estimation). 

Besides their theoretical advantage, we have shown that all quantum algorithms have demonstrated a practical quantum advantage compared to the Classical Monte Carlo method, when tested on a simulated Quantum Computer. Among them, IQAE demonstrated the most accurate estimates for a desired level of accuracy. 

The contribution of this work is to introduce the quantum algorithms which have the potential to drastically improve the Classical Monte Carlo simulations in the power systems community and test them for distributions that we often encounter in power system applications. Still, there is significant work necessary both for power system engineers and for quantum computing engineers to bring these methods into practice. Current quantum computers are at the Noisy Intermediate Scale Era. Although we attempted tests in real quantum computers, for the comparatively larger size of circuits used in the methods we presented (300 gates or more) the level of noise in real quantum computers does not allow for obtaining useful results. For the same reason, although we have shown that the number of "oracles", i.e. iterations, is indeed substantially lower in the quantum algorithms, quantum algorithms still do not run faster than Classical Monte Carlo simulations. Quantum scientists and engineers are working intensively to drastically reduce the noise level within the next few years, which will have a positive impact both in the number of gates that a quantum computer can handle and in reducing the runtime. Considering that real Quantum Computers are expected to deliver significant practical advantages very soon, we think that power system engineers shall already start developing the required methods that can address effectively power system challenges.



%

\bibliographystyle{IEEEtran}
\bibliography{References}

\begin{thebibliography}{10}
\providecommand{\url}[1]{#1}
\csname url@samestyle\endcsname
\providecommand{\newblock}{\relax}
\providecommand{\bibinfo}[2]{#2}
\providecommand{\BIBentrySTDinterwordspacing}{\spaceskip=0pt\relax}
\providecommand{\BIBentryALTinterwordstretchfactor}{4}
\providecommand{\BIBentryALTinterwordspacing}{\spaceskip=\fontdimen2\font plus
\BIBentryALTinterwordstretchfactor\fontdimen3\font minus \fontdimen4\font\relax}
\providecommand{\BIBforeignlanguage}[2]{{%
\expandafter\ifx\csname l@#1\endcsname\relax
\typeout{** WARNING: IEEEtran.bst: No hyphenation pattern has been}%
\typeout{** loaded for the language `#1'. Using the pattern for}%
\typeout{** the default language instead.}%
\else
\language=\csname l@#1\endcsname
\fi
#2}}
\providecommand{\BIBdecl}{\relax}
\BIBdecl

\bibitem{QAEBrassard}
G.~Brassard, P.~Hoyer, M.~Mosca, and A.~Tapp, ``Quantum amplitude amplification and estimation,'' \emph{AMS Contemporary Mathematics Series}, vol. 305, no.~4, June 2000.

\bibitem{Grinko_2021}
D.~Grinko, J.~Gacon, C.~Zoufal, and S.~Woerner, ``Iterative quantum amplitude estimation,'' \emph{npj Quantum Information}, vol.~7, no.~1, March 2021.

\bibitem{QAEwithoutQPE}
Y.~Suzuki, S.~Uno, R.~Raymond, T.~Tanaka, T.~Onodera, and N.~Yamamoto, ``Amplitude estimation without phase estimation,'' \emph{Quantum Information Processing}, vol.~19, January 2020.

\bibitem{Nakaji_2020}
K.~Nakaji, ``Faster amplitude estimation,'' \emph{Quantum Information and Computation}, vol.~20, no. 13{\&}14, pp. 1109--1123, November 2020.

\bibitem{yu2020comparison}
K.~Yu, H.~Lim, P.~Rao, and D.~Jin, ``Comparison of amplitude estimation algorithms by implementation,'' 2020, (unpublished).

\bibitem{emilie}
E.~Jong, ``Quantum computing for probabilistic power flow in power systems,'' Master's thesis, Technical University of Denmark, Lyngby, July 2023.

\bibitem{qiskit}
\BIBentryALTinterwordspacing
{Qiskit}, ``{An open-source framework for Quantum Computing},'' Online. [Online]. Available: \url{https://qiskit.org/}
\BIBentrySTDinterwordspacing

\bibitem{quantumriskanalysis}
S.~Woerner and D.~Egger, ``Quantum risk analysis,'' \emph{npj Quantum Information}, vol.~5, 12 2019.

\bibitem{IBMcreditrisk}
D.~J. Egger, R.~García~Gutiérrez, J.~C. Mestre, and S.~Woerner, ``Credit risk analysis using quantum computers,'' \emph{IEEE Transactions on Computers}, vol.~70, no.~12, pp. 2136--2145, 2021.

\bibitem{RAMADHANI2020106003}
\BIBentryALTinterwordspacing
U.~H. Ramadhani, M.~Shepero, J.~Munkhammar, J.~Widén, and N.~Etherden, ``Review of probabilistic load flow approaches for power distribution systems with photovoltaic generation and electric vehicle charging,'' \emph{International Journal of Electrical Power \& Energy Systems}, vol. 120, p. 106003, 2020. [Online]. Available: \url{https://www.sciencedirect.com/science/article/pii/S0142061519341730}
\BIBentrySTDinterwordspacing

\bibitem{normal1}
M.~Aien, M.~Fotuhi-Firuzabad, and F.~Aminifar, ``Probabilistic load flow in correlated uncertain environment using unscented transformation,'' \emph{IEEE Transactions on Power Systems}, vol.~27, no.~4, pp. 2233--2241, 2012.

\bibitem{plfpvev}
N.~Bhat, B.~R. Prusty, and D.~Jena, ``Cumulant-based correlated probabilistic load flow considering photovoltaic generation and electric vehicle charging demand,'' \emph{Frontiers in Energy}, vol.~11, pp. 184--196, 05 2017.

\bibitem{fee0d0cf293b443f83a0ff7b4ec7c8ac}
S.-E. Gryning, R.~Floors, A.~Pe{\~n}a, E.~Batchvarova, and B.~Br{\"u}mmer, ``\BIBforeignlanguage{English}{Weibull wind-speed distribution parameters derived from a combination of wind-lidar and tall-mast measurements over land, coastal and marine sites},'' \emph{\BIBforeignlanguage{English}{Boundary-Layer Meteorology}}, vol. 159, no.~2, pp. 329--348, 2016, {}.

\bibitem{weibullparameters}
Z.~Shu and M.~Jesson, ``Estimation of weibull parameters for wind energy analysis across the uk,'' \emph{Journal of Renewable and Sustainable Energy}, vol.~13, p. 023303, 03 2021.

\bibitem{sun2018probabilisticOPF}
W.~Sun, M.~Zamani, H.-T. Zhang, and Y.~Li, ``Probabilistic optimal power flow considering correlation of wind farms via markov chain quasi-monte carlo sampling,'' 2018.

\bibitem{PPFstochastic}
Z.~Ren, W.~Li, R.~Billinton, and W.~Yan, ``Probabilistic power flow analysis based on the stochastic response surface method,'' \emph{IEEE Transactions on Power Systems}, vol.~31, no.~3, pp. 2307--2315, 2016.

\end{thebibliography}

\end{document}